\documentclass[aps,prl,floats]{revtex4}
\normalsize
\usepackage{ifthen}
\usepackage{ifpdf}

\usepackage{latexsym}
\usepackage{amsmath}
\usepackage{amssymb}
\usepackage{bm}
\usepackage{color}

\ifpdf
\usepackage{graphicx}
\usepackage{epstopdf}
\else
\usepackage{graphicx}
\usepackage{epsfig}
\fi

\newcommand{\eexp}{\mbox{e}^}

\newcommand{\mass}{\mathsf{m}}

\newcommand{\tbox}[1]{\mbox{\tiny #1}}

\newcommand{\amatrix}[1]{\begin{matrix} #1 \end{matrix}}

\newcommand{\be}[1]{\begin{eqnarray}\ifthenelse{#1=-1}{\nonumber}{\ifthenelse{#1=0}{}{\label{e#1}}}}
\newcommand{\ee}{\end{eqnarray}}

\newcommand{\sect}[1]{{\bf #1. -- }}
\newcommand{\hide}[1]{}

\begin{document}

\title{Purely electric spin pumping in one-dimension}
\author{Yshai Avishai$^{1,3}$, Doron Cohen$^1$ and Naoto Nagaosa$^{2}$ }
\affiliation{
$^1$Department of Physics, Ben-Gurion University, Beer-Sheva 84105, Israel \\
$^2$Department of applied physics, University of Tokyo, Hongo, Bunkyo Ku, Tokyo, Japan\\
$^3$Department of physics, Hong Kong University of Science and Technology, Clear Water Bay, Kowloon, Hong Kong}

\begin{abstract}
We show theoretically that a simple one dimensional system 
(such as metallic  wire) can display quantum spin pumping 
possibly without pushing any charge. It is achieved by applying two slowly varying orthogonal gate 
electric fields on different sections of the wire, thereby generating  
local spin-orbit (Rashba) terms such that unitary transformations
at different places do not commute. 
This construction is a unique manifestation of a
spin-orbit observable effect in purely one dimensional systems
with potentials respecting time-reversal symmetry. 
\end{abstract}


\maketitle


A standard way of achieving charge transfer across a conducting system is to apply  
two gate voltages and change them adiabatically and periodically: 
Under certain conditions, a charge is transferred across the system 
during  each period. 
This is referred to as quantum (charge) pumping \cite{Thouless, BPT, Brouwer, Avron, Amnonora1}.
In recent years, the concept of pumping 
with regard to spin polarization has become a focus of attention.  
One option to get a polarized current is 
to introduce a Zeeman splitting term \cite{prop}, 
or employing ferromagnetic leads\cite{Zheng}. 
 In some cases it costs a great deal 
of dissipated energy and besides, time reversal invariance is broken. 
That motivates
the quest for achieving spin pumping without the application 
of magnetic fields \cite{Fazio, Sharma, Aono,Fu} 
(see also Ref.\cite{AmnonOra2} where spin filtering is discussed).  
It is naturally expected that pertinent experiments are 
rather difficult to carry out, and hence, an obvious desirable property 
required from a model describing spin pumping is that it should be 
simple and experimentally feasible. 

In the present work we show that spin pumping can be 
achieved in a simple one dimensional device (wire), 
by exploiting the spin-orbit (SO) interaction 
of the electron with electric fields applied 
on two different sections of the wire 
(referred below as {\em Rashba barriers}).
The model is characterized by the following attractive properties: 
{\bf (1)}~It is purely one dimensional;
{\bf (2)}~It enables pure spin (without charge) pumping;
{\bf (3)}~The expressions obtained are simple, given in analytic form;
{\bf (4)}~It serves as a pedagogical manifestation of the basic concepts of generalized forces and generalized charges;  
{\bf (5)}~It demonstrates that spin pumping is one of the few manifestations of observable SO effects in purely one-dimensional systems.

\sect{Outline}
The order of presentation is as follows: 
First we derive an expression for the scattering matrix 
of a single Rashba barrier, and then recall a composition rule 
for computing the $S$ matrix for scattering off two successive barriers. 
Once the $S$ matrix of the whole device is obtained, 
the formalism of Refs.~\cite{BPT,Brouwer} (see also \cite{pmo}) 
is employed in order to analyze the pumping process.
As a by-product an expression for the pumped spin polarization ($\vec{P}$) 
is derived, that can be regarded as an {\em SU(2)} extension 
of the Brouwer formula for the pumped charge ($Q$), 
and is somewhat simpler than the one suggested in Ref.~\cite{Sharma}.  
The conclusion includes a short summary and discussion.


\begin{figure}
\includegraphics[scale=0.5]{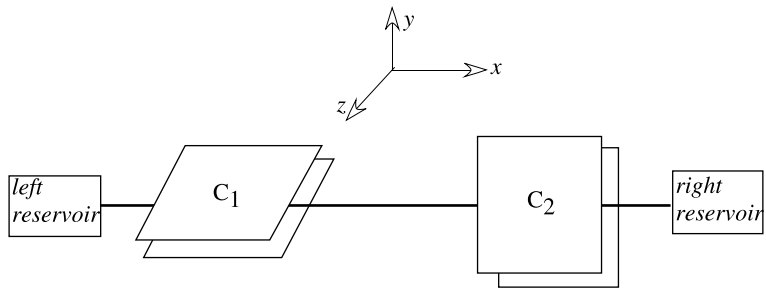}
\caption{
The pumping device (schematic). Electrons move on a one-dimensional wire 
along the $x$ direction between two reservoirs. 
Two capacitors $C_1$ and $C_2$ apply perpendicular 
electric fields ${\bm{E}_1=(0,E_1,0)}$ and  ${\bm{E}_2=(0,0,E_2)}$ 
whose strength is controlled and varied periodically by an external circuit.} 
\label{Fig1}
\end{figure}

\sect{Modeling}
The arena of our discussion is that of non-interacting electrons 
confined in a straight one-dimensional wire (along~$x$)  
possibly experiencing a scattering potential $V(x)$, 
and subject to a {\em perpendicular} electric 
field $\bm{E}(x,t)$. The Pauli Hamiltonian is 
\begin{equation} \label{standard}
\mathcal{H} =  \frac{1}{2\mass}p^2 
- \frac{e}{4\mass^2} (\bm{E} \times \bm{p} -  \bm{p} 
\times \bm{E}) \cdot \bm{\sigma}   
+ V(x) 
\end{equation} 
where $\mass$ and $e$ are the mass and the charge of the electron, 
and ${c=\hbar=1}$ units are used. 
Concretely, we have in mind  a simple and experimentally feasible example 
where the wire passes through a couple of plate capacitors $C_1$ and $C_2$ 
with different orientations, as is schematically displayed in Fig.~\ref{Fig1}.  
The fields ${\bm{E}_1(x)=(0,E_1,0)}$ and ${\bm{E}_2(x)=(0,0,E_2)}$  
are well concentrated at ${x<0}$ and ${x>0}$ segments of the wire   
and are assumed to be non overlapping.

Since the electric field is perpendicular to the wire, 
its only effect is to generate an SO 
interaction of strength $\alpha_i(x)={e E_i(x)}/{2 \mass}$ with $i=1,2$ 
corresponding to the left and right Rashba barriers.   
The dimensionless parameters that characterize 
this interaction are 
\begin{eqnarray} 
\theta_i = 2\int_{-\infty}^{\infty} \alpha_i(x') dx' 
\ \ \ \ \ \ \mbox{i=1,2} \label{theta12}
\end{eqnarray}
The time dependence of $\theta_1$ and $\theta_2$ 
is assumed to be periodic and very smooth,  
justifying the use of the adiabatic approximation. 
Practically then, the time is used as a parameter that will be employed at a later stage when 
the spin pumping is discussed (hence it will not be specified before that). 
Our first goal is to find the $\bm{S}$ matrix for scattering through 
the system depicted in the above figure. The strategy would be to write down 
the Pauli equation and solve the scattering problem 
separately for each barrier thereby obtaining the 
corresponding $\bm{S}$ matrices $\bm{S}^{(1)}$ and $\bm{S}^{(2)}$ 
and then combine them to obtain the total $\bm{S}$ matrix.

\sect{Scattering from a single Rashba barrier}
The electric field $E_1(x)$ in the left barrier is constant deep 
inside the capacitor and decays as a third power (in distance) outside it.  
For definiteness let us assume that the capacitor $C_1$ is centered 
at $x=-{L}/{2}$ and that $L$ is sufficiently large so that $E_1(x)$ 
is non-negligible only within ${-L < x < 0}$. 
The Pauli Hamiltonian Eq.(\ref{standard}) for the left barrier can be cast into the following form 
\begin{equation} \label{Pauli}
\mathcal{H}_1 = \frac{1}{2\mass}[p - \alpha_1(x)  \sigma_z]^2 + v_1(x) 
\end{equation} 
where $p$ is the momentum conjugate to $x$.  
The scalar potential ${v_1(x)=V_1(x)-\alpha_1(x)^2/2\mass}$ 
is non-vanishing within the interval ${-L < x <0}$.
The later time dependent analysis assumes that 
it is not changing in time. This assumption is legitimate 
for realistic circumstances where the $\alpha^2$ correction is small \cite{Frohlich}. 
Thus, we may start from the stationary Schr\"odinger equation for scattering 
at energy $\varepsilon$ through the first barrier, 
${\mathcal{H}_1 \psi_1(x)=\varepsilon \psi_1(x)}$, 
where $\psi_1(x)$ 
is a two component spinor.

For  the Hamiltonian Eq.(\ref{Pauli}), 
the spin projection along the $z$~direction is a `good quantum number'.
 A particle with spin up (an eigenstate of $\sigma_z$ with eigenvalue $+1$)
experiences a ``vector potential'' ${A(x)=+\alpha_1(x)}$ 
that can be {\em gauged away}, leading to an $A(x)$ 
independent reflection amplitude $r$, 
while the transmission amplitude $t$    
is multiplied by a phase factor $\eexp{i\theta_1/2}$. 
A particle with spin ``down" would experience 
a ``vector potential'' ${A(x)=-\alpha_1(x)}$    
and therefore would gain upon transmission 
an opposite phase $\eexp{-i\theta_1/2}$. 
The scattering of spin ``up" and scattering 
of spin ``down" involve different topological phases, 
turning the effect of  SO interaction to be distinct 
from that of {\em U(1)} vector potential. 
In general the scattered particle may have  
any spin direction (a superposition 
of ``up" and ``down"). Thus, due to SO interaction,  
different phases are accumulated by the up/down 
amplitudes of $\psi(x)$, implying that the 
spin direction is {\em SU(2) rotated}, 
the rotation angle being $\theta_1$.
From the above analysis it follows 
that the scattering matrix 
of the first barrier has the form  
\begin{eqnarray}
\bm{S}^{(1)}_{ab} = \left(\amatrix{
r & 0 & t\eexp{-i\theta_1/2} & 0 \cr 
0 & r & 0 & t\eexp{+i\theta_1/2} \cr
t\eexp{+i\theta_1/2} & 0 & r & 0 \cr
0 & t\eexp{-i\theta_1/2} & 0 & r} \right)
\label{S1}
\end{eqnarray}
where the channel index is  ${a=1\uparrow,1\downarrow,2\uparrow,2\downarrow}$. 
The reflection amplitude $r$ and the transmission amplitude $t$ 
are determined by the potential $v_1(x)$.       
A more compact way to write this $S$ matrix is,
\begin{eqnarray} \label{e5}
\bm{S}^{(1)} = 
\begin{pmatrix} 
R_1 & T_1' \\
T_1 & R_1 
\end{pmatrix}
=
\begin{pmatrix}
r\bm{1} & tU_1^{-1} \\
t U_1 & r\bm{1} 
\end{pmatrix},
\end{eqnarray}
where $\bm{1}$ is the $2\times2$ identity matrix 
and $U_1$ is an {\em SU(2)} rotation matrix defined via Eq.~ (\ref{S1}) 
(see also Eq.~(\ref{U12}) below). 
Within the geometry of Fig.~\ref{Fig1}, the  Hamiltonian for the second system is, 
\begin{equation} \label{Pauli2}
\mathcal{H}_2 = \frac{1}{2\mass}[p + \alpha_2(x)  \sigma_y]^2 + v_2(x). 
\end{equation} 
Assuming (just for convenience) that the second barrier has the same 
reflection and transmission amplitudes ($r$ and $t$), 
its $S$~matrix has an identical structure as $\bm{S}^{(1)}$ 
albeit with different spin rotation matrix $U_2 \ne U_1$. 
Inspecting the kinetic terms of $\mathcal{H}_1$ and $\mathcal{H}_2$ 
(see Eqs.~(\ref{Pauli}) and (\ref{Pauli2})), it is clear that the 
corresponding spin rotation matrices are,
\begin{equation} \label{U12} 
U_1=\mbox{e}^{+i \theta_1  \sigma_z/2}, \ \  U_2=\mbox{e}^{-i \theta_2  \sigma_y/2}.
\end{equation}

It is important to notice that $[U_1,U_2] \ne 0$. 
This non-commutativity of the {\em SU(2) rotations}  
is crucial for the operation of the pumping device, 
as discussed below.

\sect{Scattering from two Rashba barriers}
It is now possible to construct the  $\bm{S}=\bm{S}^{(1)} * \bm{S}^{(2)}$ 
matrix of the whole device by adding the two (non-overlaping) barriers in a series,  
employing the following prescription \cite{comb} 
for calculating the transmission and reflection amplitudes: 
\begin{eqnarray}
&& T=T_2(1-R_1^{'}R_2)^{-1}T_1, \nonumber \\
&& T'=T_1^{'}(1-R_2R_1^{'})^{-1}T_2^{'}, \nonumber \\
&& R=R_1+T_1^{'} (1-R_2R_1^{'})^{-1}R_2T_1, \nonumber \\
&& R'=R_2^{'}+T_2(1-R_1^{'}R_2)^{-1}R_1^{'}T_2^{'}.
\label{composition}
\end{eqnarray}
In the absence of SO the transmission and reflection amplitudes 
due to the total potential $v_1(x){+}v_2(x)$ are 
\begin{eqnarray}
\tau=\frac{t^{2}}{1-r^{2}}, \ \ \ \ 
\rho=r\left (1+\frac{t^{2}}{1-r^{2}} \right ). 
\label{rhotau}
\end{eqnarray}
%
In the presence of SO, Eqs.~(\ref{composition}) imply,
\begin{equation} 
\bm{S}=\begin{pmatrix} R&T'\\T&R' \end{pmatrix}=
\begin{pmatrix} 
\rho \bm{1}_{} & \tau U_1^{-1} U_2^{-1} \\ \tau  U_2 U_1 & \rho \bm{1}_{} 
\end{pmatrix},
\label{Stot}
\end{equation}
where $U_1$ and $U_2$ are defined in Eq.~(\ref{U12}).

\sect{Gauge considerations}
In one dimension, any {\em U(1)} gauge potential can be transformed away 
from the Schr\"odinger equation. 
Is it true also for the SO interaction? 
Let us introduce the transformation  
\begin{equation} 
\psi(x) = \mbox{EXP}\left[i\int_{-\infty}^x \!\!\!\!\!  dx' \bm{A}(x') \right] \Psi(x) 
\ \equiv \ U(x) \Psi(x)
\end{equation}
where $\bm{A}(x) = [\bm{\alpha}\times\bm{\sigma}]_x  = (0,-\alpha_2,\alpha_1) \cdot \bm{\sigma}$.  
The {\em EXP} stands for $x$ ordered exponentiation 
which is analogous to time ordered exponentiation. 
The result of the exponentiation is an {\em SU(2)} rotation matrix $U(x)$. 
It is not difficult to verify that $\Psi(x)$ satisfies the 
time independent equation ${\mathcal{H}_0 \Psi(x)=\varepsilon \Psi(x)}$ 
where $\mathcal{H}_0$ is obtained from Eq.(\ref{Pauli}) or Eq.(\ref{Pauli2})  after removing the $\alpha(x) \sigma$ term.
Hence, for a general barrier (not necessarily double barrier) 
the $\bm{S}$ matrix still has the structure as in Eq.(\ref{e5}) / Eq.(\ref{Stot}) 
with the appropriate rotation matrix. 

Encouraged by the above observation one may be tempted to conclude that 
the SO term can be transformed away also in the time dependent pumping formulation. 
{\em But this is wrong.} In the {\em U(1)} (charge pumping) formalism
the gauge function is ${\Lambda(x)=\int^x A(x')dx'}$
and the time dependence results in
an additional term $-d\Lambda/dt$ (electro motive force) 
that can be absorbed into the definition of the scalar potential $V(x)$.    
In the {\em SU(2)} (spin pumping) scheme the analogous transformation 
leads to a spin dependent term, and hence the SO nature of 
the interaction still manifests itself. 
Therefore in a time dependent pumping problem 
it is impossible to transform away the SO interaction.

\sect{The operation of the pumping device}
We now consider the situation displayed in Fig.~\ref{Fig1} where the 
SO dimensionless parameters  $\theta_1(t)$ and $\theta_2(t)$ of Eq.(\ref{theta12}) 
are controlled by slowly varying the fields inside the capacitors   
with a common period $2\pi/\omega$.  The adiabatic picture implies  
that the driving frequency $\omega$ is very small  
compared with other frequency scales of the system.   
Our goal is to study the pumped charge and 
the pumped spin polarization during a single period.  
The generalized conductance is defined 
as in \cite{pmo} via the relation: 
\begin{eqnarray}
dQ_a \ \ = \ \ -\sum_{i=1,2} G^{a,i}(\theta) \ d\theta_i
\end{eqnarray}
where $dQ_a$ is the charge which is pushed 
into channel~$a$ and ${i=1,2}$  
specifies the control parameter $\theta_i(t)$. 
If only one control parameter is being manipulated (call it $\theta$), 
and the practical interest is only (say) in the 
left lead, then one can use the simpler notations  
${dQ_{\uparrow} = -G^{\uparrow} d\theta}$ 
and ${dQ_{\downarrow} = - G^{\downarrow} d\theta}$. 
Accordingly the net charge which is pushed into the 
specified lead is  ${dQ = - (G^{\uparrow}+G^{\downarrow}) d\theta}$
while the net spin polarization 
is ${dP_z = - (G^{\uparrow}-G^{\downarrow}) d\theta}$. 
In what follows we explain how $dP_x$ and $dP_y$  
can be calculated as well, and show that our pump 
can generate net spin polarization current while 
the net charge current is zero at any moment.

\sect{Calculation of $G$}
The generalized conductance can be calculated  
using the Buttiker-Thomas-Pretre formula \cite{BPT,Brouwer}.  
With our notations it reads:
\begin{eqnarray}
G^{a,i}(\theta)  \ \ = \ \ 
\frac{1}{2\pi i}\left[ 
\frac {\partial \bm{S}}{\partial \theta_{i}} \bm{S}^{\dagger} \right]_{aa} 
\ \ \equiv \ \ 
-\frac{1}{2\pi}\left[\bm{H}^{(i)}\right]_{aa} 
\end{eqnarray}
If one regards the $\bm{S}(\theta)$ matrices 
as a group of unitary transformations, 
then the $\bm{H}^{(i)}$ are interpreted 
as their generators. For the problem under 
consideration:
\begin{eqnarray}
&& 
\bm{H}^{(1)}=\frac{1}{2}
\begin{pmatrix} 
|\tau|^{2} \sigma_z 
&  
\tau \rho^{*} \sigma_z U_1^{\dag} U_2^{\dag} 
\\ 
-\tau \rho^{*} U_2  U_1 \sigma_z 
& 
-|\tau|^{2} U_2 \sigma_z U_2^{\dag}  
\end{pmatrix}
\nonumber \\
&& 
\bm{H}^{(2)}=\frac{1}{2}
\begin{pmatrix} 
-|\tau|^{2} U_1^{\dag} \sigma_y U_1 
&  
-\tau \rho^{*} U_1^{\dag} U_2^{\dag} \sigma_y 
\\
\tau \rho^{*} \sigma_y U_2 U_1  
& 
|\tau|^{2} \sigma_y 
\end{pmatrix}
\label{Aexp}
\end{eqnarray}
Form the above expressions it is manifestly 
clear that the net charge which is pushed  
out into (say) the left lead is zero. 
This is because ${G^{\uparrow}+G^{\downarrow}=0}$ 
for any of the two leads.   
But what about the spin polarization current? 
The latter is determined by ${G^{\uparrow}-G^{\downarrow}}$, 
and in general it is not zero.

\sect{Pumping of spin polarization}
In order to get physical understanding of the 
spin pumping one should observe that if the channel basis 
is changed, then $\bm{H}$ undergoes a similarity transformation 
${\bm{H} \mapsto \mathcal{T}^{-1}\bm{H}\mathcal{T}}$ where $\mathcal{T}$ is the 
transformation matrix from the old to the new basis.
In particular one is interested in block diagonal $\mathcal{T}$s, 
such that each of the two $2\times2$ blocks represents 
an {\em SU(2)} rotation of the axes that are attached 
to the respective lead. 
One observes that by an appropriate choice of axes,   
a given lead-related $2\times2$ block of a given $\bm{H}$ matrix  
can be transformed into the canonical form    
\begin{eqnarray}
&& \bm{H}_{\mbox{\tiny lead} } \longmapsto 
\frac{1}{2}|\tau|^2 \sigma_{\tbox{Z}}
\end{eqnarray}
where $Z$ is the new $z$ axis.
This means that the net spin polarization 
which is pushed into a lead is 
\begin{eqnarray} \label{e18}
dP_{\tbox{Z}} =  |\tau|^2 \frac{d\theta}{2\pi}
\end{eqnarray}
where $\theta$ is either $\theta_1$ or $\theta_2$. 
It should be appreciated that the {\em direction}~($Z$)  
of the spin polarization current depends on whether 
$\theta_1$ or $\theta_2$ is being changed, 
and it is not the same for the left and for the right lead. 
Specifically, if $\theta_1$ is being varied, 
then the spin polarization of the current in the left lead
is in the $z$~direction,
while in the right lead it is in the $xy$ plane, 
with an angle $\theta_1$ relative to the $y$~direction.

\sect{Example}
As an example one can consider the 
following prototype pumping cycle 
\begin{eqnarray}
(\theta_1,\theta_2)=(0,0)\mapsto (\pi,0) \mapsto (\pi,\pi) \mapsto (0,\pi) \mapsto ...
\end{eqnarray}
During the 1st and the 3rd stages of the cycle 
the spin polarization which is pushed into 
the left lead is $\pm|\tau|^2/2$ in the $z$ 
direction, while during the 2nd and 4th stages 
it is $+|\tau|^2/2$ in the $y$ direction.  
Thus we generate per cycle net spin polarization $|\tau|^2$ in the~$y$ direction, 
while at any moment the net pumped charge is zero.

\sect{{\em SU(2)} extension of the Brouwer formula}
For a general pumping cycle the net pumping is 
given by a line integral over the conductance.  
Following Brouwer, one can replace this line integral 
by an area integral using Stokes theorem. Namely,  
\begin{eqnarray}
Q_a \ \ = \ \ \oint \bm{G} \wedge d\bm{\theta} \ \ = \ \ \iint \bm{C}_{aa} d\theta_1d\theta_2
\end{eqnarray}
where $\bm{\theta}=(\theta_1,\theta_2)$,   
and $\bm{G}=(G^{a,1},G^{a,2})$. 
In the above expression we have introduced 
the ``rotor" $\bm{C}_{aa}$ of $\bm{G}$. This rotor 
can be regarded as a diagonal element of the 
matrix ${\bm{C} = -(1/2\pi)[\partial_1 \bm{H}_2 - \partial_2 \bm{H}_1}] $, hence       
\begin{eqnarray}
\bm{C} \ = \ \frac{1}{\pi} \Im \left[\left (\frac {\partial \bm{S}}{\partial \theta_{2}} \right ) 
\left (\frac {\partial \bm{S}^{\dagger}}{\partial \theta_{1}} \right )\right] 
 = \frac{1}{2\pi i} [\bm{H}_2,\bm{H}_1]
\end{eqnarray}
Note that if one change the channel basis, 
then $\bm{C}$ undergoes a similarity transformation.
Calculating $\bm{C}$ for our model system one observes 
that the derivatives bring down $\sigma_y \sigma_z=i \sigma_x$ 
and each $\sigma_x$ is rotated by the corresponding {\em SU(2)} 
rotation matrix, by angles $-\theta_1$ around $z$ for $U_1$ 
and $\theta_2$ around $y$ for $U_2$, leading to  
\begin{eqnarray} \label{Bexp}
&& {\bf C}
= \frac{i}{4\pi} |\tau|^{2} 
\begin{pmatrix} 
U_1^{\dag}\sigma_{y} \sigma_{z} U_1 & 0 \\
0 & U_2 \sigma_{y} \sigma_{z} U_2^{\dag} 
\end{pmatrix}  \\
&& 
= -\frac{1}{4\pi}|\tau|^{2}
\begin{pmatrix} 
\cos\theta_{1} \sigma_{x}+ \sin\theta_{1} \sigma_{y} & 0 \\
0 & \cos\theta_{2} \sigma_{x}-\sin\theta_{2} \sigma_{z} 
\end{pmatrix} \nonumber 
\end{eqnarray} 
One can re-write the expressions for the pumped charge 
and the pumped spin polarization in a Brouwer-like style:
\begin{eqnarray} 
Q_{\mbox{\tiny lead}} \ \ &=& \ \ \iint \mbox{trace} (\bm{C} \bm{1}{\mbox{\tiny lead}} )  
d\theta_1d\theta_2
\\
\label{pol}
\vec{P}_{\mbox{\tiny lead}} \ \ &=& \ \ \iint \mbox{trace} 
(\bm{C} \vec{\bm{\sigma}}{\mbox{\tiny lead}} )  d\theta_1d\theta_2
\end{eqnarray}
The matrix $\bm{1}{\mbox{\tiny lead}}$ is a projector on (say) the left lead, 
which means in practical terms that one can keep 
only the upper right $2\times2$ block of~$\bm{C}$, 
and sum only over the channels of the left lead. 
For the model system under consideration we manifestly 
have zero trace and hence the Brouwer formula gives $Q_{\mbox{\tiny lead}}=0$. 
This is to be expected when the effect of spin orbit appears 
as a pure gauge: it affects the wave function merely through an {\em SU(2)} 
phase factor. On the other hand, spin pumping is not zero 
because ${\bf C}$ is multiplied by spin matrices before being traced, 
and one can get a non-zero spin polarization $\vec{P}$.

%
It might be useful to notice that the integral over $\bm{C}$,  
which is Eq.(\ref{pol}) without the trace, 
is formally an expression for the spin polarization matrix,   
rather than for the spin polarization vector:
\begin{eqnarray} 
\rho_{\mbox{\tiny lead}} \ \ = \ \ \iint \bm{C}_{\mbox{\tiny lead}}   d\theta_1d\theta_2
\end{eqnarray}
Here $\bm{C}_{\mbox{\tiny lead}}$ is the relevant $2\times2$ block that 
corresponds to the lead under consideration.

\sect{Discussion}
On the practical level it has been demonstrated in this work 
that it is possible to polarize a neutral spin current 
using a strictly 1D device with no extra magnetic fields. 
This should be contrasted with more complicated arrangements 
that were suggested for this purpose e.g. in Ref.\cite{AmnonOra2}. 
The scheme that has been considered in our analysis 
is based on a pumping (time dependent) paradigm, 
instead of the conventional transmission filter paradigm, 
and at the same time does not involve the use of magnetic fields.     

On the mathematical side an extremely simple result 
for the pumped spin polarization has been obtained, 
namely  Eq.(\ref{e18}). As demonstrated, 
it can also be formulated as an {\em SU(2)} extension 
of the Brouwer formula for charge pumping, noting that 
the geometric (Kubo) conductance $G$ is formally 
a 2-form (curvature), while $C$ is a 3-form (scalar).   

We have illuminated the gauge consideration in the theory:
while in the time-independent setting it is possible 
to transform away the SO interaction, in spite of the 
non-commutativity of the {\em SU(2)} gauge transformations,  
this is no longer true for the time dependent Hamiltonian  
that describes the pumping scenario.

\sect{Acknowledgments}
The research of D.C is partially supported by a grant from the USA-Israel
Binational Science Foundation (BSF), and by a grant 
from the Deutsch-Israelische Projektkooperation (DIP). That of Y.A is partially supported by an ISF grant. 
N.N. is supported by Grant-in-Aids  under Grant No.\ 19048015, 21244053, and
NAREGI Nanoscience Project from the Ministry of Education, Culture, Sports, Science
and Technology, Japan.


\end{document}